\documentclass[a4paper,11pt]{article}
\usepackage{pos}

\title{QCD Equation of State and Phase Diagram from Holographic Black Holes}
\author*[a]{Joaquin Grefa}
\author[b]{Jorge Noronha}
\author[b]{Jacquelyn Noronha-Hostler}
\author[a]{Israel Portillo}
\author[a]{Claudia Ratti}
\author[c]{Romulo Rougemont}

\affiliation[a]{Physics Department, University of Houston,\\
  Houston TX 77204, USA}

\affiliation[b]{Illinois Center for Advanced Studies of the Universe, Department of Physics, University of Illinois at Urbana-Champaign,\\
Urbana, IL 61801, USA}

\affiliation[c]{Departamento de F\'{i}sica Te\'{o}rica, Universidade do Estado do Rio de Janeiro,\\
Rua S\~{a}o Francisco Xavier 524, 20550-013, Maracan\~{a}, Rio de Janeiro, Rio de Janeiro, Brazil}

\emailAdd{vgrefaju@central.uh.edu}
\emailAdd{jn0508@illinois.edu}
\emailAdd{jnorhos@illinois.edu}
\emailAdd{iportill@central.uh.edu}
\emailAdd{cratti@central.uh.edu}
\emailAdd{romulo.pereira@uerj.br}

\abstract{By using the AdS/CFT correspondence, we construct an Einstein-Maxwell-Dilaton model to map the thermodynamics of strongly interacting matter. The holographic model, constrained to reproduce the lattice QCD equation of state at zero baryon chemical potential, predicts a critical end point and a first order phase transition line. We also obtain the equation of state of the model for a large region of the phase diagram. We characterize the crossover transition region by two lines, one associated to the inflection of the second order baryon susceptibility, and another one associated to the minimum of the square of the speed of sound along trajectories of constant entropy per baryon number. We observe that both lines merge at the critical point.}

\FullConference{%
	The International conference on Critical Point and Onset of Deconfinement - CPOD2021\\
  	15 – 19 March 2021\\
 	Online - zoom
 }

\begin{document}
\maketitle

\section{Introduction}
The QCD phase diagram, usually represented in a plane of temperature and baryon chemical potential, is the subject of active research in the theory of strong interactions. Lattice QCD calculations at vanishing chemical potential have shown that strongly interacting matter undergoes a rapid crossover transition \cite{Aoki:2006we} characterized by a change in degrees of freedom from the low temperature hadrons to a novel high temperature phase of matter corresponding to a deconfined state of quarks and gluons, i.e. the quark-gluon plasma (QGP). The extreme conditions needed for this phenomenon to take place can be reproduced in relativistic heavy ion collisions at the Large Hadron Collider at CERN, and the Relativistic Heavy Ion Collider at BNL and have shown the QGP is the smallest and hottest fluid that exhibits the smallest viscosity over entropy density ratio $\eta/s$ ever observed \cite{Heinz:2013th}.

Additionally, it is conjectured that the crossover at zero chemical potential evolves into a line of first order phase transition with a critical end point at increasing baryon chemical potential, as suggested by the effects of nonzero quark mass in chiral models \cite{Stephanov:1998dy}. It is the goal of the experiments to find the QCD critical point by systematically decreasing the center of mass energy in relativistic heavy ion collisions, thus increasing the baryon chemical potential and scanning the phase diagram. Unfortunately, lattice QCD calculations cannot be performed at finite chemical potentials due to the Fermi sign problem, even though there have been efforts to circumvent this issue such as the Taylor expansion of the pressure and baryon density for small chemical potential. As a consequence, most of the QCD phase diagram remains unexplored. Therefore, another model is needed to guide the experimental search for the predicted critical point and the line of first order phase transition at increasing chemical potential.
Such a model should reproduce the equation of state from lattice QCD at small baryon chemical potential, and exhibit the near perfect fluidity of the QGP implied by experimental data from heavy ion collisions. \cite{Critelli:2017oub}. The present work summarizes the results of a holographic model \cite{Grefa:2021qvt} that fulfils such requirements.

\section{The Holographic Model}
The Einstein-Dilaton-Maxwell (EMD) model is described by the simplest, 5-dimensional gravitational action that can generate a QCD-like theory, \cite{DeWolfe:2011ts,Critelli:2017oub,Grefa:2021qvt},
\begin{equation}
   S= \frac{1}{2\kappa_{5}^{2}}\int_{\mathcal{M}_5} d^{5}x\sqrt{-g}\left[R-\frac{(\partial_\mu \phi)^2}{2}-V(\phi)-\frac{f(\phi)F_{\mu\nu}^{2}}{4}\right],
\end{equation}
where $g_{\mu\nu}$ and $\kappa_{5}^{2}=8\pi G_{5}$ are the 5-dimensional metric tensor and gravitational constant respectively, $R$ is the Ricci scalar, and $\phi$ is the dilaton field.
Since QCD is a nonconformal quantum field theory, the dynamical breaking of the conformal symmetry is driven by a scalar potential of the dilaton field $V(\phi)$, which is a free function in the holographic model and determines the thermodynamics of the QCD-like theory at zero chemical potential. The dilaton potential $V(\phi)$ is fixed by solving the equations of motion and constraining the holographic equation of state to match the corresponding lattice QCD results shown in \cite{Borsanyi:2013bia}. Additionally, effects due to a finite baryon chemical potential can be taken into account by introducing a Maxwell field $A_{\mu}$ in $F_{\mu\nu}=\partial_{\mu}A_{\nu}-\partial_{\nu}A_{\mu}$ with another free function $f(\phi)$ that couples the Maxwell and dilaton fields. This coupling function is fixed by matching the holographic second order baryon susceptibility $\chi_{2}^{B}$ to the corresponding lattice result at $\mu_{B}=0$ in \cite{Bellwied:2015lba}. Additionally, by taking the AdS radius $L$ to be unity, an energy scale $\Lambda$ can be introduced to convert the physical observables from holographic units to physical ones in powers of MeV \cite{Critelli:2017oub,Grefa:2021qvt}.

The metric for charged black holes, spatially isotropic and translationally invariant, can be described by the following Ansatz \cite{DeWolfe:2011ts},
\begin{equation}
    ds^2 = e^{2A(r)}[-h(r)dt^2+d\vec{x}^2]+\frac{dr^{2}}{h(r)},
\end{equation}
which also considers that the dilaton field is a function of the holographic direction $\phi=\phi(r)$ as well as the Maxwell field $A_{\mu}dx^{\mu}=\Phi(r)dt$. The black hole fields, $\phi(r),h(r),A(r),\Phi(r)$, near horizon Taylor coefficients needed for the numerical integration of the equations of motion can be parametrized by two initial conditions $(\phi_{0},\Phi_{1})$ which are the value of the dilaton field and the electric field at the horizon \cite{DeWolfe:2011ts,Critelli:2017oub}. With the ultraviolet coefficients of the black hole fields, it is possible to obtain the temperature, $T$, baryon chemical potential, $\mu_{B}$, entropy density, $s$, and baryon density, $\rho_{B}$, for the QCD-like theory by making use of the holographic dictionary, that is, a pair of initial conditions $(\phi_{0},\Phi_{1})$ results into a mapping of the state variables $(s,\rho_{B})$ into a plane of temperature $T$, and baryon chemical potential $\mu_{B}$. 

\section{Results}
\begin{figure}[h]
     \centering
   \includegraphics[width=0.49\textwidth]{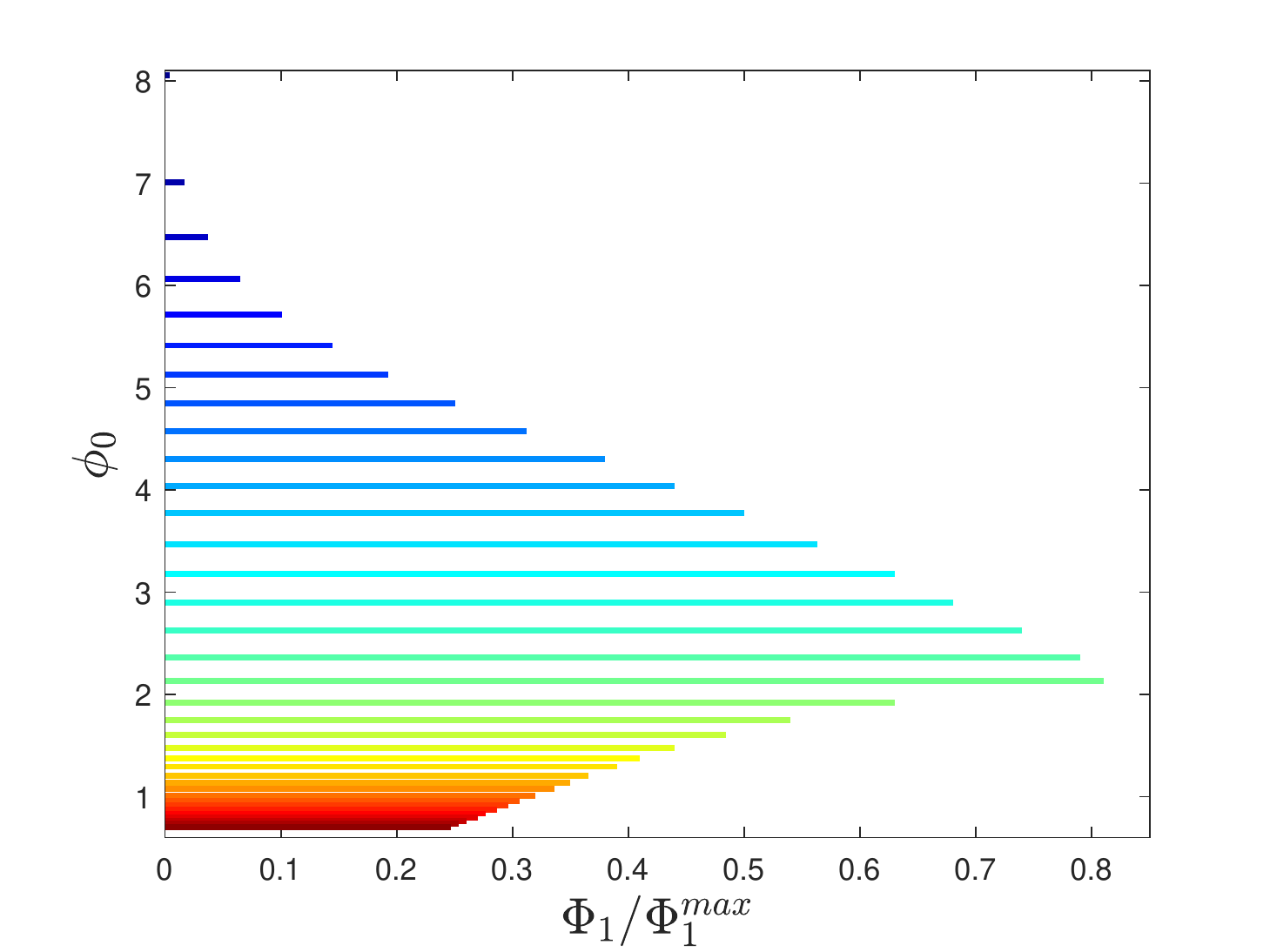}
   \includegraphics[width=0.49\textwidth]{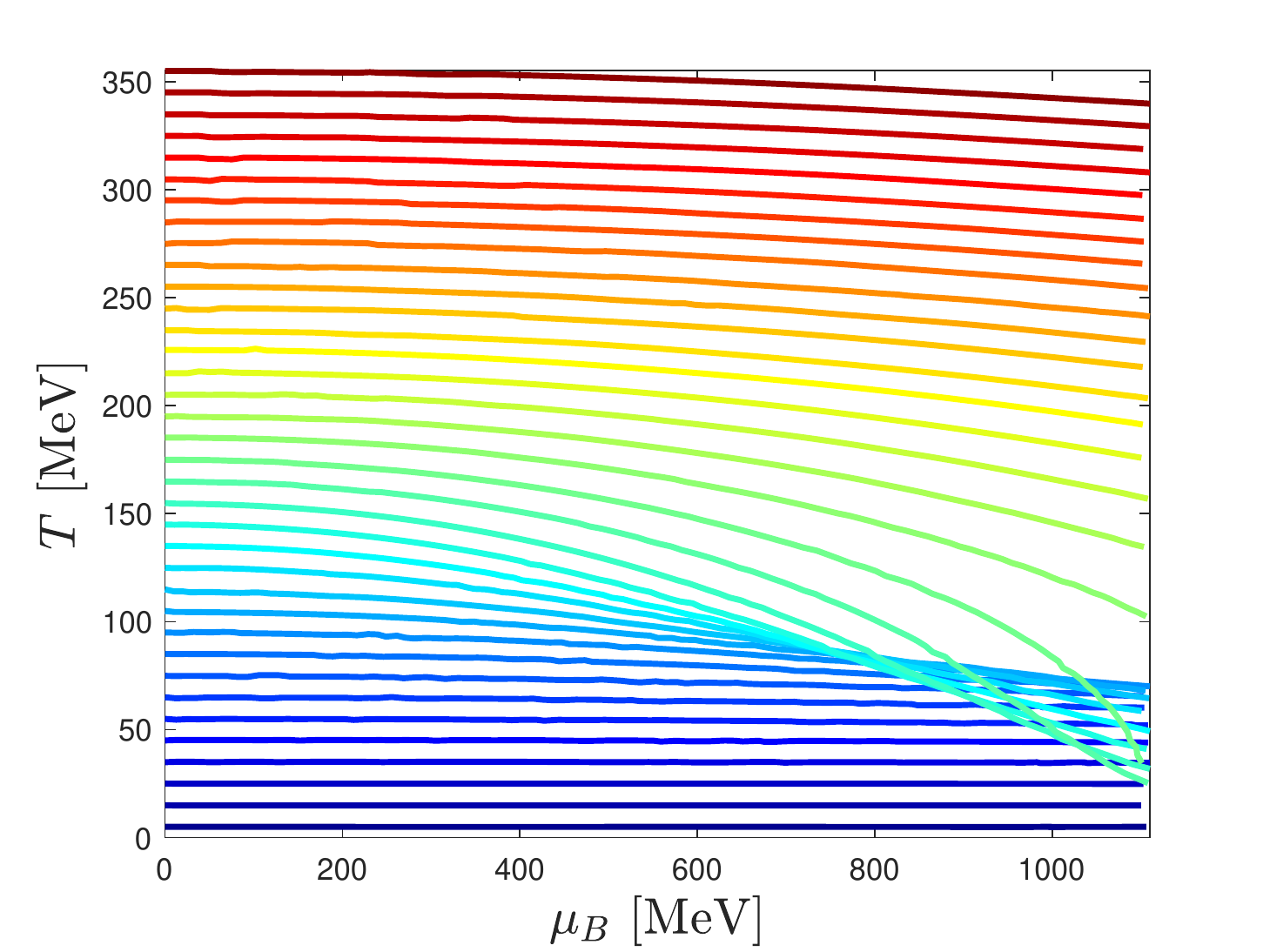}  
    \caption{Mapping of the black hole initial conditions into a rectangular region of the QCD phase diagram.}
    \label{fig:mapping2}
\end{figure}

An example of how the black hole initial conditions $(\phi_{0},\Phi_{1})$ should be taken to map a rectangular region in the QCD phase diagram is shown in Fig.\ \ref{fig:mapping2}. The value of $\phi_{0}$ is roughly translated as the temperature $T$, and provides the EoS at zero chemical potential when $\Phi_{1}=0$. On the other hand, a nonzero $\Phi_{1}$ is mapped into a finite baryon chemical potential value. The lines in Fig. \ref{fig:mapping2} are produced by first obtaining equally spaced values of temperature when varying $\phi_{0}$ at $\Phi_{1}=0$. The lines are color coded to easily notice that large values of $\phi_{0}$ produce small values of temperature and  vice versa. Then, for each value of $\phi_{0}$, $\Phi_{1}$ is varied starting from zero to cover a region up to $\mu_{B}=1100$ MeV, although it could be increased up to its maximum value where the solutions to the black hole fields are still asymptotically AdS. The lines shown in Fig. \ref{fig:mapping2} clearly exhibit a region of three competing solutions in the QCD phase diagram which suggest the presence of a critical point where this region starts. The exact point where the overlapping solutions begin $(T_{c}=89 \textrm{ MeV},\mu_{B}^{c}=724 \textrm{ MeV})$ was identified as the critical point in this current holographic model by finding the location in the holographic phase diagram where the second order baryon susceptibility $\chi_{2}^{B}$ diverges \cite{Critelli:2017oub}.

\begin{figure*}[hbt!]
    \centering
    \includegraphics[width=0.7\textwidth]{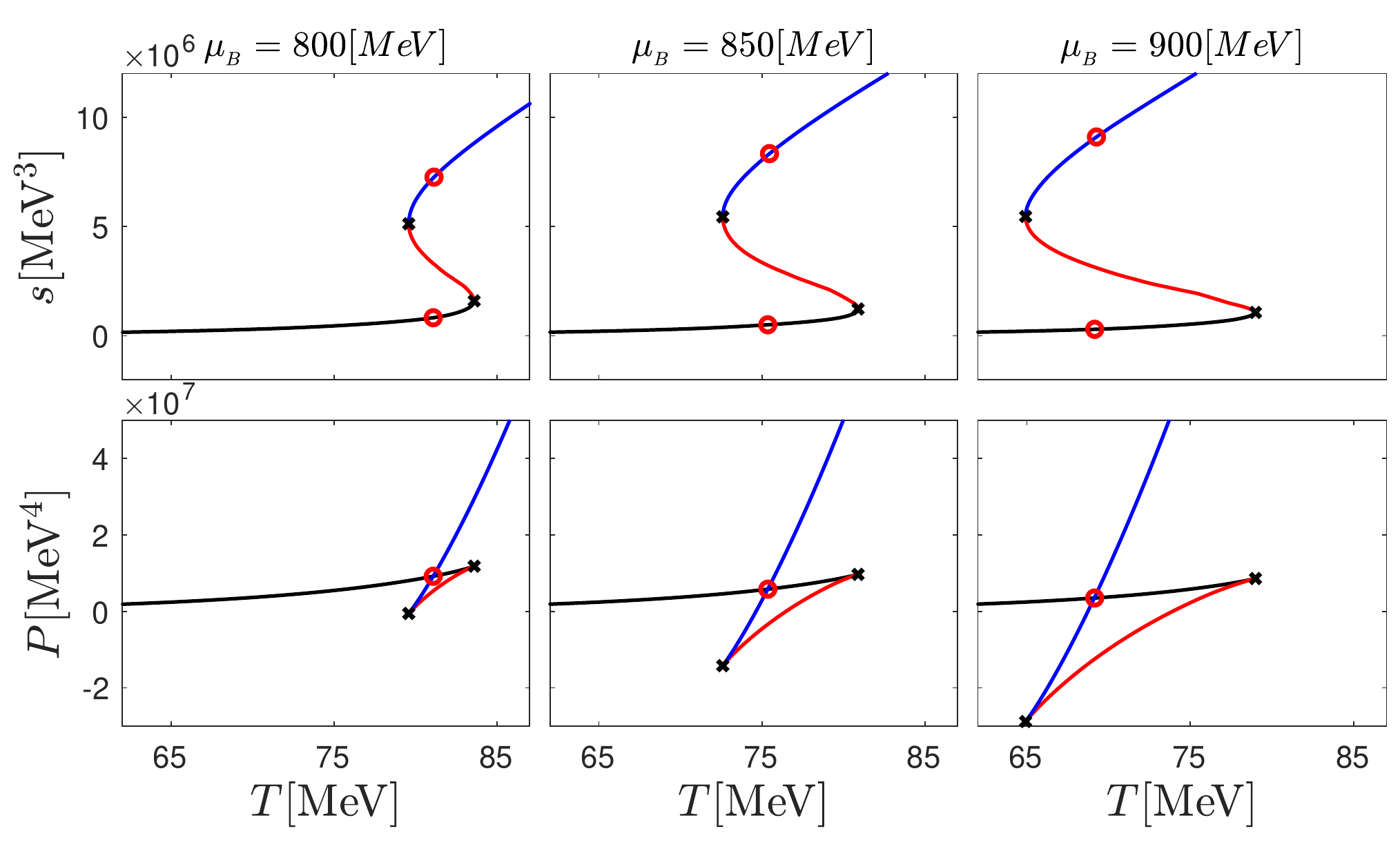}
    \caption{Entropy density $s$ (upper panels) and its integral with respect to the temperature, corresponding to the pressure (lower panels), for three different values of $\mu_{B} > \mu_{B}^{\textrm{CEP}}\sim 724$ MeV.}
    \label{fig:multivalued_entropy}
\end{figure*}

Beyond the critical point, the solutions of the black hole EOM generates the coexistence region that characterizes a first order phase transition and corresponds to all the extrema of the free energy although only one of these solutions minimizes the free energy. To the right of the critical point, i.e. $T<T_{c}$ and $\mu_{B}>\mu_{B}^{c}$, the entropy and baryon density exhibit a characteristic \textit{S} shape  which means that at a given temperature we can find three different values for these observables. The case for the entropy density $s$ is shown in Fig. \ref{fig:multivalued_entropy}. In order to obtain the first order phase transition line, the entropy density was integrated with respect to the temperature over the multivalued region, and we located the point where the resulting curve, which corresponds to the pressure or minus the free energy, crosses itself. This procedure is analogous to a Maxwell equal-area construction, although it is computationally easier to perform. The result of this procedure is shown in the lower panel of Fig. \ref{fig:multivalued_entropy}.

Because of the nature of the crossover, from vanishing chemical potential up to the critical point, there is not a unique method to characterize a transition temperature. On the contrary, a transition temperature can be obtained from the inflection point or the extrema of quantities sensitive to a change of degrees of freedom from the confined hadronic phase to a system of deconfined quarks and gluons \cite{Borsanyi:2010bp,Grefa:2021qvt}. For this holographic model, we have chosen to characterize the crossover by computing the speed of sound squared at constant entropy per baryon number $c_{s}^{2}$, and the inflexion point of the second order baryon susceptibility $\chi_{2}^{B}$. In the case of the square of the speed of sound, a formula that considers only derivatives of the pressure in the direction of the chemical potential or the temperature was used since it is practical over a regular grid, and reads \cite{Parotto:2018pwx,Floerchinger:2015efa}, 
\begin{equation}\label{eq:c2s}
c_{s}^{2}=\frac{\rho_{B}^{2}\partial_{T}^{2}P-2s\rho_{B}\partial_{T}\partial_{\mu_{B}}P +s^{2}\partial_{\mu_{B}}^{2}P}{(\epsilon+P)[\partial_{T}^{2}P\partial_{\mu_{B}}^{2}P-(\partial_{T}\partial_{\mu_{B}}P)^{2}]}.
\end{equation}

\begin{figure}[h]
     \centering
   \includegraphics[width=0.49\textwidth]{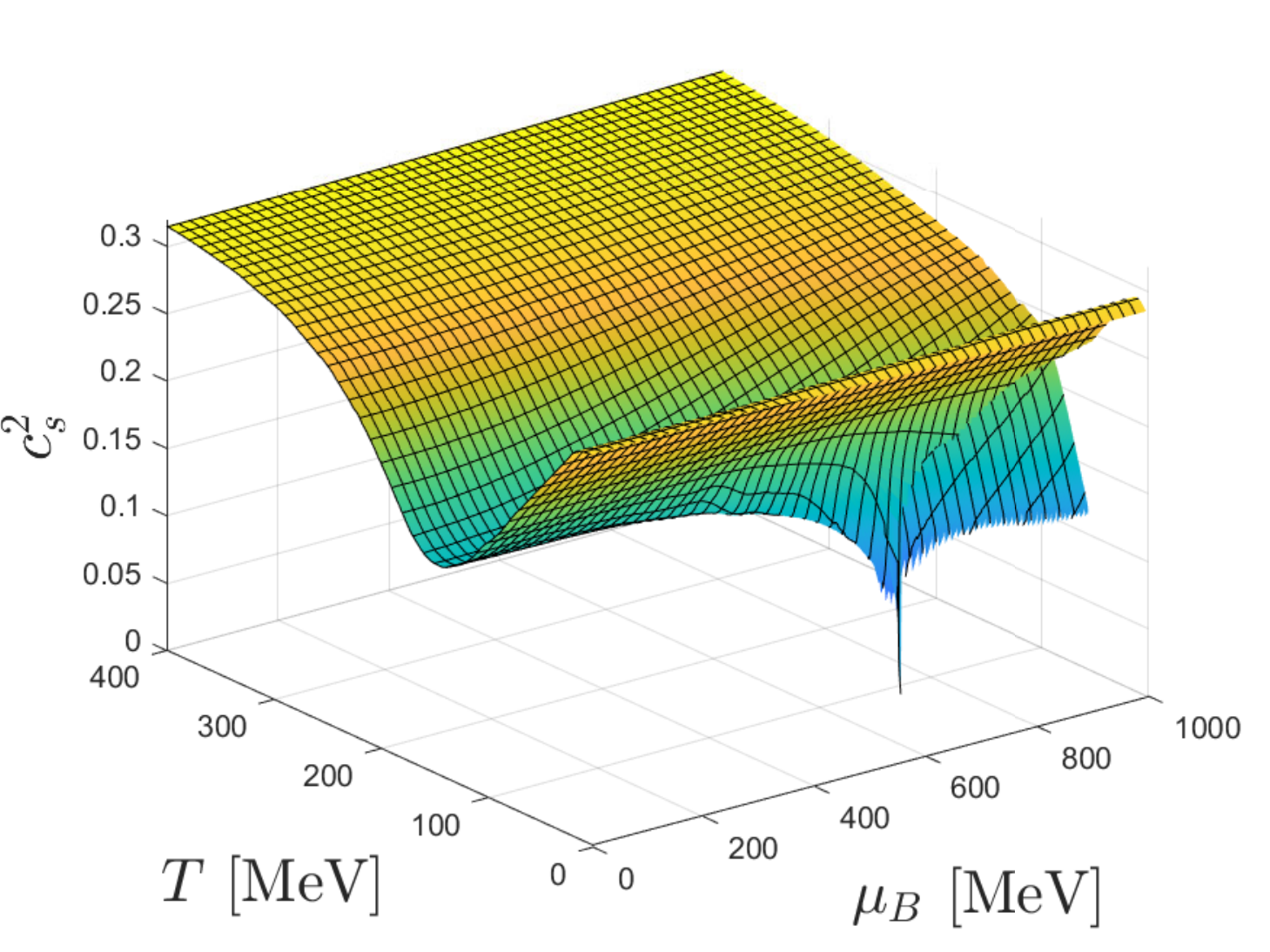}
   \includegraphics[width=0.49\textwidth]{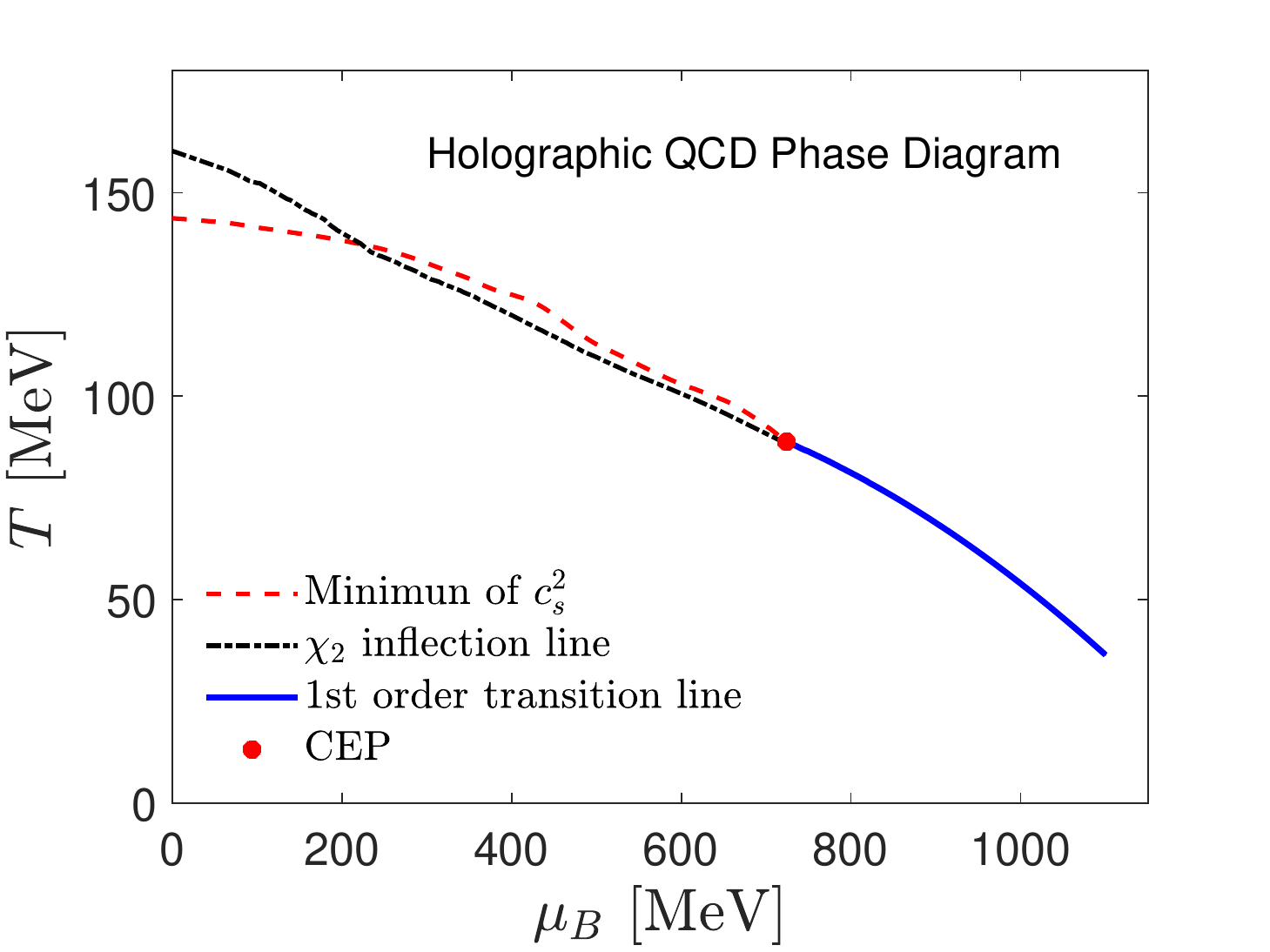}  
    \caption{Left: the speed of sound squared at constant entropy per baryon number computed from Eq. \ref{eq:c2s}. Right: the phase diagram of our EMD model. The inflection point of $\chi_2^{B}$ and the minimum of $c_{s}^{2}$  are used to characterize the crossover region.}
    \label{fig:phase_diagram}
\end{figure}

However, since the derivatives of the pressure diverge at the critical point, and the numerical noise associated to the derivatives increases near the critical point, it was necessary to replace the noisy points with the result from computing the derivative of the pressure with respect to the energy density $c_{s}^{2}=(\partial P/\partial \epsilon)$ over several isentropic trajectories and use an interpolation method to cover the regions affected by the noise. The result of the mentioned procedure is shown in Fig. \ref{fig:phase_diagram}. The inflection point of $\chi_2^{B}$ and the minimum of $c_{s}^{2}$ converge to the location of the critical point where the line of first order phase transition starts.

\section{Conclusions}

The holographic model is constrained to mimic the lattice QCD equation of state and second order susceptibility to study the behaviour of strongly interacting matter at finite baryon chemical potential. The QCD phase diagram can be mapped from a pair of initial conditions $(\phi_{0},\Phi_{1})$ for the black hole fields to obtain the entropy density $s$ and baryon density $\rho_{B}$ in a plane of temperature and baryon chemical potential for a QCD-like theory.

The crossover, described in this work by the inflection point of the second order baryon susceptibility and the minimum of the square of the speed of sound at constant entropy per baryon number, evolves into a first order phase transition line with a critical point characterized by the divergence of the second order baryon susceptibility or the global minimum of the speed of sound squared at the critical point located at $T_{c}=89$ MeV and $\mu_{B}^{c}=724$ MeV. The transition line is found by integrating the entropy density $s$ with respect to the temperature over the multivalued region and obtaining the points where the resulting curve intersects itself.

%\section*{Acknowledgements}
\acknowledgments
This material is based upon work supported by the National Science Foundation under grants no. PHY-1654219, PHY-2116686 and OAC-2103680 and by the US-DOE Nuclear Science Grant No. DE-SC0020633, US-DOE Office of Science, Office of Nuclear Physics, within the framework of the Beam Energy Scan Topical (BEST) Collaboration. J.N. is partially supported by the U.S. Department of Energy, Office of Science, Office for Nuclear Physics under Award No. DE-SC0021301. R.R. acknowledges financial support by Universidade do Estado do Rio de Janeiro (UERJ) and Fundação Carlos Chagas de Amparo à Pesquisa do Estado do Rio de Janeiro (FAPERJ).
%\begin{thebibliography}{99}
%\bibitem{...}
%....
%\end{thebibliography}
%\bibliographystyle{unsrt}
\bibliographystyle{JHEP}
\bibliography{references}

\end{document}